\documentclass[usenatbib]{mn2e}
\usepackage{amsmath, amssymb}
\usepackage{amsfonts}
\usepackage{epsfig,rotating,subfigure}

\def\simeq{
\mathrel{\raise.3ex\hbox{$\sim$}\mkern-14mu\lower0.4ex\hbox{$-$}}
}

\def\ltsima{$\; \buildrel < \over \sim \;$}
\def\simlt{\lower.5ex\hbox{\ltsima}}
\def\gtsima{$\; \buildrel > \over \sim \;$}
\def\simgt{\lower.5ex\hbox{\gtsima}}

\def\lsun{{\rm L_{\odot}}}
\def\msun{{\rm M_{\odot}}}

\def\be{\begin{equation}}
\def\ee{\end{equation}}

\def\del#1{{}}

\newcommand{\apj}{ApJ}

\newcommand{\mnras}{MNRAS}
\newcommand{\aap}{A\&A}
\newcommand{\araa}{ARA\&A}
\newcommand{\apjl}{ApJL}

\title{Galaxy--wide outflows: cold gas and star formation at high speeds}

\author[Kastytis~Zubovas, Andrew~R.~King]{Kastytis~Zubovas$^{1}$ and Andrew~R.~King$^2$\\
$^1$ Center for Physical Sciences and Technology, Savanori\c{u} 231, Vilnius LT-02300, Lithuania\\
$^2$ Theoretical Astrophysics Group, University of Leicester, Leicester LE1 7RH, U.K.\\
{E-mail:~} {\rm kastytis.zubovas@ftmc.lt}}

\begin{document}

\maketitle

\begin{abstract}

Several active galaxies show strong evidence for fast ($v_{\rm out}
\sim 1000~{\rm km\,s}^{-1}$) massive ($\dot{M} =$ several $\times
1000~\msun\,{\rm yr}^{-1}$) gas outflows. Such outflows are expected
on theoretical grounds once the central supermassive black hole
reaches the mass set by the $M - \sigma$ relation, and may be what
makes galaxies become red and dead. Despite their high velocities,
which imply temperatures far above those necessary for molecule
dissociation, the outflows contain large amounts of molecular gas. To
understand this surprising result, we investigate the gas cooling and
show that the outflows cannot stably persist in high--temperature
states. Instead the outflowing gas forms a two--phase medium, with
cold dense molecular clumps mixed with hot tenuous gas, as
observed. We also show that efficient cooling leads to star formation,
providing an observable outflow signature. The central parts of the
outflows can be intrinsically luminous gamma--ray sources, provided
that the central black hole is still strongly accreting. We note also
that these outflows can persist for $\sim 10^8$~yr after the central
AGN has turned off, so that many observed outflows (particularly with
high speeds) otherwise assumed to be driven by starbursts might also
be of this type.

\end{abstract}


\section{Introduction}

Over the past decade, observations revealed mounting evidence of
strong outflows from active galactic nuclei (AGN), which significantly
affect the evolution of their host galaxies. Highly ionized
relativistic winds \citep[e.g.][]{Pounds2003MNRASa, Pounds2003MNRASb,
  PoundsVaughan11} and kiloparsec-scale outflows
\citep{Feruglio2010A&A, Sturm2011ApJ, Rupke2011ApJ} have been detected
in active galaxies, with total kinetic luminosities $L_{\rm kin}$
equal to a few percent of the AGN luminosity.

The basic properties of large scale outflows -- mass and momentum flow
rate, velocity $v_{\rm out} \sim 1000$~km/s, kinetic energy -- can be
well explained by the model of AGN wind feedback, first proposed by
\citet{King2003ApJ} and later developed both analytically
\citep{King2005ApJ,King2010MNRASa,Zubovas2012ApJ,Faucher2012MNRASb}
and numerically \citep{Nayakshin2010MNRAS,Zubovas2012MNRASa}. Within
this model, AGN radiation pressure launches a relativistic wind from
very close in, where the Thomson scattering optical depth
self-regulates to be $\tau=1$. The wind then shocks against the
surrounding gas and drives an outflow. If the mass of the supermassive
black hole (SMBH) that powers the AGN exceeds the critical mass given
by
\begin{equation}\label{eq:msigma}
M_\sigma \simeq \frac{f_{\rm c} \kappa \sigma^4}{\pi G^2} \simeq 3.67 \times
10^8 \sigma_{200}^4 \; \msun,
\end{equation}
where $f_{\rm c} = 0.16$ is the cosmological ratio of gas density to
matter density, $\kappa \simeq 0.4$~cm$^2$/g is the electron
scattering opacity and $\sigma \equiv 200 \sigma_{200}$~km/s is the
velocity dispersion in the host galaxy spheroid, its wind shocks can
propagate out to large distances and are no longer efficiently Compton
cooled. As a result, the previously weak and cold, momentum--driven
outflows change character to become energy--driven. These outflows are
far more violent, and clear galaxies of gas \citep{Zubovas2012ApJ},
leaving them red and dead.

The observed large-scale outflows contain mostly molecular gas. This
is, in princple, a problem for the AGN wind feedback model, because
outflowing gas is accelerated by a shock and heated to temperatures of
$10^6-10^7$~K, much higher than molecular dissociation
temperatures. Although cold clumps could be embedded in the ambient
medium and carried with the flow, such acceleration would produce
signatures incompatible with observations \citep{Cicone2012A&A}.

We show here that radiative cooling rapidly puts a large fraction of
the outflowing material into molecular form, which stably coexists
with tenuous hot gas. This simultaneously explains the large velocity
of the molecular component and its immunity to high Mach number
shocks, in agreement with observations \citep{Cicone2012A&A}. We
review the properties of the forward shock that the outflow drives
into the ISM in Section \ref{sec:shock} and show that it cannot be
kept thermally stable at high temperatures by quasar radiation in
Section \ref{sec:stability}. We then consider the cooling of its gas
in Section \ref{sec:cooling} and estimate the rate of star formation
in the cooling outflow in Section \ref{sec:starform}. Finally, we
discuss our results in Section \ref{sec:discussion} and summarize
giving predictions for observations in Section \ref{sec:summary}.

\section{The outer shock} \label{sec:shock}

\begin{figure}
  \centering
    \includegraphics[width=0.49\textwidth]{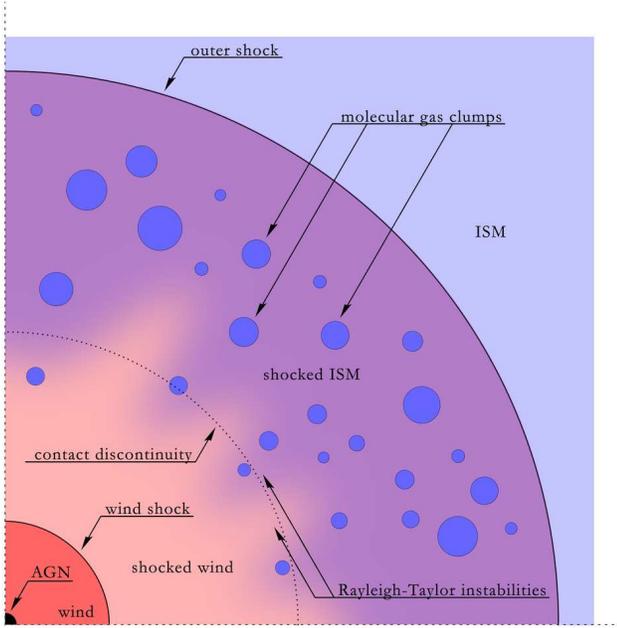}
  \caption{A schematic view of the multiphase nature of the
    energy--driven outflow launched when the central supermassive
    black hole reaches the critical $M - \sigma$ mass.  The AGN drives
    a powerful quasi-spherical wind with speed $v \sim 0.1c$ from its
    accretion disc. This wind is strongly shocked just outside the
    inner Compton cooling radius (radius $R_{\rm IC} \sim
    0.5M_8/\sigma_{200}$~kpc, where $M_8$ is the black hole mass in
    units of $10^8\msun$). The expanding shocked wind sweeps up and
    drives an outer shock into the host ISM. The contact discontinuity
    between the expanding wind gas and swept--up ISM is
    Rayleigh--Taylor unstable, so these two components mix together in
    the outflow and help maintain a constant pressure. The outflow
    cools radiatively, and most of it freezes out into clumps of cold
    molecular material. Despite its high velocity $v \sim 1000~{\rm
      km\, s^{-1}}$ most of the outflow is in molecular form. The very
    high temperature of the wind shock near $R_{\rm IC}$ implies that
    these outflows should be intrinsically luminous gamma--ray
    sources.}
  \label{fig:cartoon}
\end{figure}

The dynamics of spherically-symmetric energy--driven outflows from
accreting AGN are described in \citet{King2005ApJ} and
\citet{Zubovas2012ApJ}. We briefly summarize this subject here, to
establish the properties of the outflow relevant for the present
problem.

Rapidly accreting AGN launch fast ($v_{\rm w} \sim 0.1 c$) winds from
their accretion discs. The wind shocks against the surrounding ISM and
drives an outflow. Close to the black hole, the shocked wind cools
rapidly by the inverse Compton effect of the AGN radiation field, and
the resulting outflow of swept--up gas is relatively slow and must
ultimately fall back. But once the mass of the central black hole
grows above the critical value $M_\sigma$ (eq. \ref{eq:msigma}), the
wind shock and swept--up ISM can reach large radii, where the AGN
radiation field is too dilute to cool the shocked wind gas. Now all of
the wind energy goes into powering the outflow, rather than almost all
being lost to cooling. This extra energy accelerates the outflow,
increasing its velocity at the contact discontinuity between the wind
and the interstellar medium to
\begin{equation} \label{eq:ve}
v_{\rm e} = \left(\frac{2 \eta\sigma^2c}{3 f}\right)^{1/3} \simeq 925
\sigma_{200}^{2/3} f^{-1/3} {\rm km s}^{-1},
\end{equation}
where $\eta = 0.1$ is the radiative efficiency of accretion and $f
\equiv f_{\rm g}/f_{\rm c}$ is a factor allowing for deviations of the
gas-to-total density ratio $f_{\rm g}$ from the cosmological value
\citep[see][for derivation of $v_{\rm e}$]{King2005ApJ}. An outer
shock moves ahead of this discontinuity into the ISM, sweeping the
latter up and compressing it. Adiabatic jump conditions give a
velocity $v_{\rm out} = 4/3 v_{\rm e}$ for this forward shock
\citep{Zubovas2012ApJ}. This heats the ISM to a temperature
\begin{equation} \label{eq:temp}
T_{\rm sh} = \frac{3}{16}\frac{\mu m_{\rm p} v_{\rm out}^2}{k} \simeq 2.2
\times 10^7 \sigma_{200}^{4/3} f^{-2/3} {\rm K},
\end{equation}
where $k$ is Boltzmann's constant and $\mu m_{\rm p}$ is the mean mass per
particle. The particle density of the swept--up ISM is
\begin{equation}\label{eq:nshock}
n = \frac{2 f_{\rm g} \sigma^2}{\pi G \mu m_{\rm p} R^2} \simeq 60 f
\sigma_{200}^2 R_{\rm kpc}^{-2} {\rm cm}^{-3},
\end{equation}
which is just $4$ times the preshock particle density, as appropriate
for a strong shock. 

The interstellar gas which has passed through the outer shock is
relatively dense, and so likely to cool quite rapidly below the shock
temperature. The primary cooling process at the shock temperature is
free-free cooling, and outflows are optically thin to this radiation:
\begin{equation}\label{eq:taushock}
\tau \simeq \kappa n \mu m_{\rm p} R_{\rm sh} \simeq 0.2 f \kappa
\sigma_{200}^2 R_{\rm kpc}^{-1},
\end{equation}
where we take $\mu = 0.63$, appropriate for fully ionized gas of Solar
metallicity. The pressure within the shocked outflow is approximately
constant \citep{Zubovas2013MNRASb}. Furthermore, the interface between
the shocked wind and the outflow is highly Rayleigh-Taylor (RT)
unstable \citep[also see Section
  \ref{sec:stability}]{King2010MNRASb}. RT fingers mix the shocked
cooling interstellar gas with the dilute hot wind gas inside it,
maintaining a constant pressure within the whole outflow. This
pressure decreases over time as $t^{-2}$.

From the above considerations, the following picture of a composite
outflow emerges. Most of the outflowing mass is in the form of the
shocked ISM at temperatures below a few times $10^7$~K. This gas is
mixed (by RT instabilities) and maintains a pressure equilibrium with
the much hotter shocked wind. We show this schematically in Figure
\ref{fig:cartoon}. To determine the observational appearance of this
mixture we need to consider whether the shocked ISM can be kept at
these large temperatures for a long time.

\section{Outflow stability} \label{sec:stability}

An important aspect of the thermal evolution of outflowing gas is its
phase structure. The gas can either cool as a single-phase medium and
experience strong compression by the shocked wind into a narrow shell,
or it can develop a two-phase structure with cold gas embedded in the
tenuous, high-pressure hot medium. Therefore we consider whether a
multiphase equilibrium between cold gas clumps and a surrounding
diffuse outflow is possible.

Following \citet{Krolik1981ApJ}, we define a pressure-based
ionization parameter
\begin{equation} \label{eq:ion}
\Xi = \frac{L}{4 \pi R^2 n k c T} = \frac{p_{\rm rad}}{p_{\rm gas}},
\end{equation}
where $L$ is the quasar luminosity, which we take to be a fraction $l$
of the Eddington luminosity, with the mass of the SMBH a fraction $m$
of the critical $M_\sigma$ mass (see Section \ref{sec:shock}), so that
\begin{equation}
L = \frac{4 \pi G M c }{\kappa} l = \frac{4 f_{\rm c} \sigma^4 c}{G} m l.
\end{equation}
Substituting the expressions for luminosity, density
(eq. \ref{eq:nshock}) and temperature (eq. \ref{eq:temp}) into
eq. (\ref{eq:ion}), we find
\begin{equation}
\Xi = 0.07 f^{-1/3} m l,
\end{equation}
where we expect $l \simeq m \simeq 1$ in general.

Figures 4 and 5 of \citet{Krolik1981ApJ} show that for $\Xi \lesssim
0.5$ the outflowing gas can only be stable in a cold phase. The
ionization parameter does not change as the outflow expands, because
both radiation and gas pressures decrease as $p_{\rm rad} \propto
p_{\rm gas} \propto R^{-2}$. Therefore, the only way to establish an
equilbrium is to reduce the gas fraction $f$. As the gas cools and
molecular clumps detach from the hot flow, the density in the latter
decreases, but a reduction by a factor $\sim 400$ is necessary before
equilibrium can be established. Therefore we predict that the diffuse
gas envelope surrounding the molecular clumps is very tenuous, and
almost all the outflow mass is in molecular form.

The separation of the outflow into a two-phase medium is not a new
result. \citet{King2010MNRASb} showed that energy-driven outflows are
highly RT unstable, therefore a complicated mixture of shocked wind
and outflow gas would emerge even without any cooling
processes. \citet{Nayakshin2012MNRASb} used numerical simulations to
show that cooling outflows fragment, but did not investigate the
reasons for fragmentation in detail. Gravitational instability of
expanding supernova shells was predicted analytically by
\citet{Whitworth1994A&A}.

\section{Cooling of the shocked gas} \label{sec:cooling}

\begin{figure}
  \centering
    \includegraphics[width=0.49\textwidth]{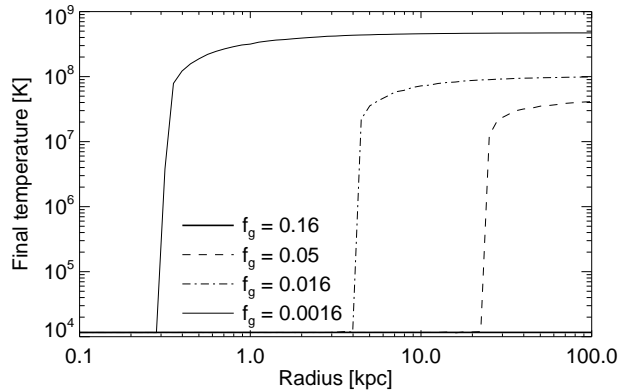}
  \caption{Temperature of shocked interstellar gas one dynamical time
    after the shock heats it, as a function of radius $R$. For low gas
    fractions and at large radii, radiative two-body cooling is
    inefficient, so the gas stays hot ($T > 10^7$~K). There is a sharp
    transition at a particular radius for every gas fraction; within
    this radius, gas is cool ($< 10^5$~K) and the resulting two--phase
    instability leads to the formation of molecules. The line at $1.1
    \times 10^4$~K is a lower limit of our integrator rather than a
    physical barrier.}
  \label{fig:outcool}
\end{figure}

\begin{figure}
  \centering
    \includegraphics[width=0.49\textwidth]{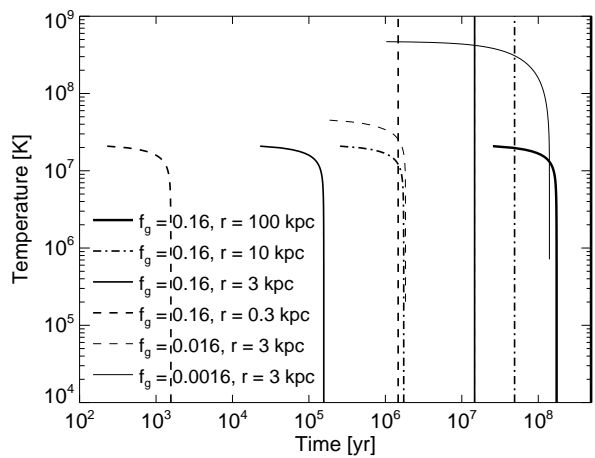}
  \caption{Cooling of pressure-confined gas in an outflow as function
    of time, for a range of gas fractions and outer shock radii. The
    gas is assumed to be stationary (assumption appropriate since
    cooling is faster than dynamical time). The cooling process is
    slow at first, but then accelerates rapidly as the gas begins to
    contract and Compton heating becomes inefficient. Vertical lines
    show dynamical timescales at certain radii (same line style as
    corresponding cooling curves of $f_{\rm g} = 0.16$).}
  \label{fig:outtemp}
\end{figure}

We now turn our attention to the efficiency of cooling processes
acting in the shocked ISM. The outer shock temperature is
approximately equal to the Compton temperature ($T_{\rm C} \simeq 1.9
\times 10^7$~K), so Compton cooling -- the primary process of cooling
the shocked wind -- does not affect the shocked ISM. Instead, at
temperatures $T_{\rm g} > 10^4$~K, gas predominantly cools via
two--body processes, i.e. bremsstrahlung and metal line cooling. To
estimate the cooling timescale, we numerically integrate the cooling
function from \citet[see their Appendix C for
  details]{Sazonov2005MNRAS}, which is appropriate for optically thin
gas illuminated by quasar radiation; as we showed above
(eq. \ref{eq:taushock}), this is a valid assumption. We perform the
calculations for a range in gas fraction, $f_{\rm g}$, between $1.6
\times 10^{-3}$ and $1$, and at distances from $R = 100$~pc to
$100$~kpc. For each of those, we integrate the cooling for one
dynamical time $\sim R/\sigma$, assuming the pressure is constant and
equal to that of the shocked wind gas.

The results are plotted in Figures \ref{fig:outcool} and
\ref{fig:outtemp}. The first one shows the gas temperature after one
dynamical time as a function of radius for several gas fractions
$f_{\rm g}$. There is a clear difference between two regimes: low
density gas far away from the SMBH cools inefficiently, while closer
in or at higher density, cooling rapidly decreases gas temperature to
the lower limit of the adopted cooling function, $T_{\rm low} \simeq
1.1 \times 10^4$~K.  A sharp transition seen in the low density curves
($f_{\rm g} \leq 0.05$) can be identified as a cooling radius similar
to the cooling radius \citep{Zubovas2012MNRASb} of the wind shock. From
the integration results, we find this outer cooling radius to be
\begin{equation}
R_{\rm OC} \simeq 100 f^{1.5} {\rm kpc},
\end{equation}
where `OC' stands for `outflow cooling'. Therefore, as long as the gas
density is $\simgt 0.05$ times the total density, the shocked
interstellar gas in the central few kiloparsecs is efficiently
cooled. As it cools, the gas is likely to fragment into cold clumps
surrounded by a hot diffuse medium. A precise treatment of this
fragmentation is beyond the scope of this Letter; we only note that
\citet{Nayakshin2012MNRASb} showed the existence of such fragmentation
numerically.

\subsection{Molecular cooling processes} \label{sec:molcool}

Once the gas temperature falls below a few times $10^4$~K, atomic and
molecular hydrogen and metal cooling become important. Calculating the
effect of these elements precisely is very complicated, as cooling
rates strongly depend on gas opacity, chemical abundance and various
other local parameters. Here we only make a rough estimate. We take
the cooling rate of atomic gas just below $10^4$~K to be
\begin{equation}
\Lambda_{\rm mol} \simeq 10^{-25} n'^2~{\rm erg\,cm}^3\,{\rm s^{-1}},
\end{equation}
as given in \citet[Chapter 6.2]{Spitzer1978book}. This is a
conservative estimate, as it does not include the contributions of
metal lines or molecular hydrogen. The gas density appropriate for
this cooling process is $n' \sim n T_{\rm sh} / T_{\rm low} \simeq
10^3 n$, because the clumps of atomic gas are much denser than the
average density in the shocked ISM. Substituting eq. (\ref{eq:nshock})
gives the cooling rate
\begin{equation}
\Lambda_{\rm mol} = 3.6 \times 10^{-16} R_{\rm kpc}^{-4} f^2
\left(\frac{T_{\rm sh} / T_{\rm low}}{10^3}\right)^2 \sigma_{200}^4~{\rm
  erg}\,{\rm s^{-1}}.
\end{equation}
The corresponding cooling time, assuming the rate is constant, is
\begin{equation}
t_{\rm cool, mol} \simeq \frac{3 n' k T_{\rm low}}{\Lambda} \simeq 24
\left(\frac{T}{1.1 \times 10^4 {\rm K}}\right)^2 R_{\rm kpc}^2 f^{-1/3}
\sigma_{200}^{-10/3}~{\rm yr}.
\end{equation}
This cooling timescale is much shorter than dynamical time for any
reasonable value of $R$. We conclude that once gas cools down to
atomic temperatures by two--body processes, the subsequent atomic and
molecular cooling is very efficient, and rapidly puts most of the
outflowing matter into molecular form.

\section{Star formation in the outflow} \label{sec:starform}

Cold clumps of molecular gas, possibly shielded from the quasar
radiation by envelopes of atomic hydrogen (the warm phase of the ISM),
are perfect locations for star formation. As a result, we expect
large-scale AGN outflows to form stars moving with large radial
velocities. Such an effect was found in numerical simulations by
\citet{Nayakshin2012MNRASb} and \citet{Zubovas2013MNRAS}, but they
used only rough approximations for the rate of outflow
fragmentation. Here we provide four estimates for an upper limit of
star formation rate in the AGN outflow:

a) The cooling timescale estimate. This simply gives the rate at which
gas in the outflow is converted into the cold molecular phase. The
cooling timescale can be approximated by a power-law
\begin{equation}
t_{\rm cool} \simeq 0.017 R_{\rm kpc}^2 f^{-1.75} \; {\rm Myr}.
\end{equation}
The total mass of the outflowing gas is
\begin{equation}
M_{\rm out} \simeq 3 \times 10^9 R_{\rm kpc} f \sigma_{200}^2 \;
\msun.
\end{equation}
Therefore, the approximate gas cooling rate is
\begin{equation}
\dot{M}_{\rm cool} \simeq 1.8 \times 10^5 R_{\rm kpc}^{-1} f^{2.75}
\sigma_{200}^2 \; \msun {\rm yr}^{-1}.
\end{equation}
This estimate, however, assumes that there is enough hot gas consumed
by the outflow to be cooled down. This is not the case in reality,
since the cooling timescale is typically faster than the flow
timescale.

b) The ``locally dynamical'' estimate. Independently of how rapidly
the gas cools, it cannot fragment faster than on its local dynamical
timescale. For gas of density $n' = 10^3 n$, this timescale is
\begin{equation}
t_{\rm dyn} \simeq 0.13 R_{\rm kpc} f^{-1/2} \sigma_{200}^{-1} {\rm
  Myr},
\end{equation}
which translates into a star formation rate of
\begin{equation}
\dot{M}_{\rm dyn,local} \simeq 2.4 \times 10^4 f^{3/2} \sigma_{200}^2
\; \msun {\rm yr}^{-1}.
\end{equation}

c) The true parameter governing the rate of gas cooling and
fragmentation is the rate at which the outflow sweeps up new mass:
\begin{equation} \label{eq:sfr_dyn}
\dot{M}_{\rm out} \simeq 3750 f^{2/3} \sigma_{200}^{8/3} \; \msun {\rm
  yr}^{-1}.
\end{equation}
These three limits scale differently with galaxy and outflow
parameters, so the true dynamical limit is the lowest of the
three. However, this is only the limit on gas cooling and
fragmentation rate, because it does not account for star formation
feedback.

d) A rough estimate of the true star formation rate can be made by
assuming that star formation feedback (in the form of photoionization,
massive star winds and supernova explosions) heats and disperses the
cold gas. In this way, the star formation process self-regulates so
that the feedback energy injected by the stars balances the cooling
rate of the gas. The energy loss rate is
\begin{equation}
\dot{E}_{\rm cool} \sim \frac{M_{\rm out} k T_{\rm sh}}{\mu m_{\rm p}
  t_{\rm cool}} \sim 3.2 \times 10^{46} R_{\rm kpc}^{-1} f^{2.08}
\sigma_{200}^{10/3} \; {\rm erg s}^{-1}.
\end{equation}
The stellar feedback luminosity that affects the gas is
\begin{equation}
L_{\rm fb} \sim \epsilon_{\rm f} \epsilon_* \dot{M}_* c^2,
\end{equation}
where $\epsilon_{\rm f} < 1$ is a feedback coupling efficiency and
$\epsilon_* = 7 \times 10^{-3}$ is the stellar mass-to-luminosity
conversion efficiency \citep{Leitherer1992ApJ}. Requiring that the
feedback luminosity balances the cooling rate gives
\begin{equation} \label{eq:sfr_en}
\dot{M}_* \simeq 80\epsilon_{\rm f}^{-1} R_{\rm kpc}^{-1}
f^{2.08} \sigma_{200}^{10/3} \; \msun {\rm yr}^{-1}.
\end{equation}
Since not all of the stellar luminosity is absorbed by the surrounding
material (due to both geometry and varying opacity), the true upper
limit to the star formation rate is somewhat higher.

The above estimate does not depend on the rate at which the outflow
accumulates mass. If all the gas is cold, star formation still
self-regulates at a value set by balancing the actual cooling of gas
(heated by the stellar feedback processes rather than the forward
shock) with the energy input rate.

This star formation rate results in a luminosity of young stars of
$L_* \simeq \dot{E}_{\rm cool}/\epsilon_{\rm f} \simeq 8.3 \times
10^{12} \epsilon_{\rm f}^{-1} R_{\rm kpc}^{-1} f^{2.08}
\sigma_{200}^{10/3} \; \lsun$, comparable to or slightly higher than
the luminosity of the AGN that drives the outflow.

\subsection{Star formation efficiency}

We make a rough estimate of the typical size of gravitationally bound
fragments. Since the cooling timescale is much shorter than dynamical
for most values of $R$ and $f$, the outflowing gas cools and fragments
in a very narrow region, defined by the cooling timescale $t_{\rm
  cool}$ and the sound speed in the post-shock gas $c_{\rm s,hot}
\simeq v_{\rm out}$:
\begin{equation} \label{eq:dcool}
d_{\rm cool} \sim v_{\rm out} t_{\rm cool} \simeq 21 R_{\rm kpc}^2
f^{2.08} \sigma_{200}^{2/3} \; {\rm pc}.
\end{equation}
If we assume that the clumps form from material initially distributed
in regions with length scales of order $d_{\rm cool}$, the clump mass
is of order
\begin{equation}\label{eq:clumpmass}
M_{\rm clump} \sim d_{\rm cool}^3 \mu m_{\rm p} n \sim 8600 R_{\rm kpc}^4
f^{7.25} \sigma_{200}^4 \; \msun,
\end{equation}
i.e. the clumps have masses typical of molecular clouds. The high
outflow pressure ensures that their densities are much larger than
typical for molecular clouds \citep{Roman-Duval2010ApJ}. For the same
reason, their radii are also significantly smaller than $d_{\rm
  cool}$.

Under normal conditions, the star formation efficiency per dynamical
time in clouds of this mass is $\sim 0.02$
\citep{McKee2007ARA&A}. However, the star formation rate is probably
somewhat higher here due to the high external pressure; in principle,
the situation is the same as that of an AGN outflow compressing
pre-existing dense structures in the galaxy, such as the galactic disc
\citep{Zubovas2013MNRASb}. If we consider the estimate based on energy
balance (eq. \ref{eq:sfr_en}) as the true limit to the star formation
rate, the efficiency then becomes
\begin{equation}
\epsilon_{\rm SF} \simeq \frac{\dot{M}_*}{\dot{M}_{\rm out}} \simeq
0.02 \epsilon_{\rm f}^{-1} R_{\rm kpc}^{-1} f^{1.42}
\sigma_{200}^{2/3},
\end{equation}
which is $> 0.02$ since $\epsilon_{\rm f} < 1$. On scales of
individual clumps, the star formation efficiency might be
significantly higher due to their large density
\citep{Kruijssen2012MNRAS}. In this case, it would be larger cooling
complexes that are affected and dispersed by self-regulating feedback.

\section{Discussion} \label{sec:discussion}

The picture described here suggests that most of the gas in a
spherically symmetric energy--driven AGN outflow is not thermally
stable and cools to low temperatures as the outflow proceeds. Both
two--body (bremsstrahlung and metal line) and atomic hydrogen cooling
processes are very efficient and cool the gas to temperatures
conducive to molecule formation. This happens on a timescale $t_{\rm
  cool} \simeq 0.017 R_{\rm kpc}^2 f^{-1.75}$ Myr, which is much lower
than the dynamical timescale within a radius $R_{\rm OC} \simeq 100
f^{1.5}$~kpc, where $f$ is the ratio of gas density in the galaxy to
the cosmic baryon density. This cool component is mixed in with much
hotter tenuous gas from the shocked central wind, and partially
entrained by it.

This picture compares favourably with observations of fast
molecular outflows \citep{Feruglio2010A&A,Sturm2011ApJ,Rupke2011ApJ}.
\citet{Zubovas2012ApJ} already showed that the predicted velocities
($\sim 1000~{\rm km\,s}^{-1}$) and mass outflow rates (several
$1000~\msun\,{\rm yr}^{-1}$) are in broad agreement with expectations
from the energy--driven phase triggered by the central SMBH mass
reaching the critical $M -\sigma$ value (eq. \ref{eq:msigma}). Our
results here refine this picture: we show that most of the outflow
appears in molecular form, despite its high velocity $v \sim 1000~{\rm
  km\, s^{-1}}$. Once again, this is consistent with observations of
outflows \citep{Cicone2012A&A}.

\subsection{Dynamics of molecular clumps}

The clumps of molecular gas have much higher densities than their
surroundings and are less affected by the ram pressure of the AGN
wind. As a result, they partially detach from the diffuse outflow and
decelerate. \citet{Zubovas2013MNRAS} found the same result in
numerical simulations of fragmenting outflows. In addition,
Rayleigh-Taylor instabilities mean that even the diffuse shocked ISM
does not expand uniformly, but instead has a spread of velocities
below the formal outflow velocity calculated by
\citet{Zubovas2012ApJ}. This explains the fact that observed
velocities of molecular outflows are typically lower than the formal
$v_{\rm out} = (4/3) v_{\rm e}$ (see eq. \ref{eq:ve}).

\subsection{Molecular gas distribution}

As the outflow progresses, on average, regions closer to the black
hole are denser and cooler than regions further away. This happens due
to two reasons: the cooling rate decreases outwards, and the outer
regions are shocked later, leaving them less time to cool down. Such a
situation is consistent with observations of \citet{Cicone2012A&A},
which suggest that at least in the molecular outflow of Mrk 231,
molecular gas has higher densities in the central regions than in the
outskirts.

It is important to note that these calculations assume a spherically
symmetric isothermal gas distribution in the galaxy. In a realistic
galaxy, this assumption breaks down outside a few kpc from the
centre. While this does not preclude the formation of molecular gas in
the outflow, the cooling rate may be significantly reduced in the
directions perpendicular to the galactic disc plane. In this case, the
molecular outflow would appear more similar to a galactic fountain,
with most clumps ejected close to the plane of the disc. The star
formation rate in the outflow would be similarly reduced, and the
existing star formation may be easily confused with star formation in
the disc due to its location.

\subsection{Structure and stability of the outflow} \label{sec:stability_discussion}

\citet{King2010MNRASb} showed that in energy-driven outflows, the
density contrast between the shocked wind and the shocked ambient
medium is of order $10^9$. This configuration leads to formation of
strong RT instabilities in the contact discontinuity, which enhance
mixing and maintain a constant pressure throughout the outflow (see
Section \ref{sec:shock}).

Several authors found that RT instabilities may be suppressed under
certain conditions. \citet{Mizuta2005ApJ} showed that recombination
effectively suppresses the instability growth, at least on length
scales $\sim 0.1-1$~pc. However, efficient recombination requires
temperatures $<10^5$~K and so is irrelevant for the gas immediately
behind the shock. Instabilities with wavelengths shorter than the
width of the cooling region, i.e. $\lambda \simlt 20$~pc (see
eq. \ref{eq:dcool}) are not suppressed by recombination effects.

Another way of suppressing RT instabilities is by strong radiation
pressure \citep{Jiang2013ApJ}. However, as we showed in Section
\ref{sec:stability}, the ratio of radiation pressure to gas pressure
is $\ll 1$ until almost all of the outflowing gas becomes
molecular. Therefore, radiation pressure is also unable to stop RT
instability from growing.

There may be other ways of stabilising the contact discontinuity, for
example via magnetic fields \citep[e.g.,][]{Jones2005ApJ}. But even in
this case, short wavelength instabilities can grow
\citep{Vishniac1983ApJ}, leading to mixing across the
discontinuity. Finally, once the shocked outflowing gas becomes
gravitationally unstable, the resulting clumpiness allows some of the
hot wind to leak out of the bubble, once again enhancing mixing.

The net effect of these instabilities is to maintain pressure
equilibrium throughout the outflow. Even without the instabilities,
the pressure varies only by a factor of a few from the contact
discontinuity to the forward shock \citep{Zubovas2013MNRASb}, so the
precise nature of the instabilities is not important to the overall
picture of outflow cooling. Another important effect of RT fingers is
that they produce a large number of shock fronts, which can accelerate
particles to cosmic ray (CR) energies. Our model therefore predicts
that CRs are produced in AGN unless all instabilities of the contact
discontinuity are efficiently suppressed.

\subsection{Survival of molecular clumps}

Molecular clumps embedded in hot gas are subject to various processes
that act to destroy them. Here we briefly review the effects of three
processes: cloud evaporation, Rayleigh-Taylor instabilities and
Kelvin-Helmholtz instabilities.

The timescale of cloud evaporation can be approximated using the
expressions in \citet{Cowie1977ApJ}. Their equation (22), when
rescaled appropriately, gives
\begin{equation}
t_{\rm evap} \sim 4 \times 10^7 M_4 r_{\rm pc}^{-1} T_7^{-5/2} {\rm
  yr} \sim 6 \times 10^6 R_{\rm kpc}^2 f^{6.83} {\rm yr},
\end{equation}
where $M_4$ is clump mass in $10^4 \; \msun$ and $r_{\rm pc}$ its
radius in parsecs. In the above expression, we used the clump mass
estimate from eq. (\ref{eq:clumpmass}), the hot gas temperature from
eq. (\ref{eq:temp}) and estimated the radius from the assumption that
the clump density is $10^3$ times the particle density
(eq. \ref{eq:nshock}). It appears that in most situations, i.e. where
$R_{\rm kpc} > 1$ and $f \simeq 1$, large clumps survive for long
enough to form a significant number of stars. However, the clumps
should have a range of masses around the calue given by
eq. (\ref{eq:clumpmass}), and the smaller clumps can evaporate more
rapidly. This echoes the numerical simulations of
\citet{Marcolini2005MNRAS}, who find clouds with masses of order $200
\; \msun$ evaporating due to thermal conduction over $t \sim 1$~Myr
under similar conditions, although with the wind streaming past the
clouds with large velocities.

Another aspect which increases the survival time of clumps in our
model is their rapid radial motion. The evaporation timescale scales
with the square of the distance, so as the clump moves outward, it is
progressively less affected by evaporation.

Rayleigh-Taylor instabilities can in principle develop at the clump
interface, since the clump is much denser than the wind. Several
authors have found, however, that in hot environments, hydrogen
recombination \citep{Ricotti2014MNRAS} and thermal conduction
\citep{Marcolini2005MNRAS} suppress the growth of these
instabilities. Therefore we are confident the RT instabilities do not
destroy the clumps rapidly enough to prevent them from being observed
at large distances.

Finally, Kelvin-Helmholtz instabilities can develop and disrupt the
clumps, but only if the shear velocity between the clump and the
diffuse outflow gas is large. Within our model, clumps form from the
fast-moving gas, and therefore move with roughly the same velocity,
i.e. $v_{\rm rel} = \left|v_{\rm out} - v_{\rm cl}\right| \ll v_{\rm
  out}$, so the KH instability growth timescale $t_{\rm KH} \sim
\sigma/v_{\rm rel} t_{\rm dyn} \gg t_{\rm dyn}$.

Our discussion here is brief, since a more detailed treatment of clump
dynamics after formation is beyond the scope of the
paper. Nevertheless, even if some process destroys the clump soon
after formation, the clump gas only adds to the hot outflow. The hot
gas then cools down and forms clumps again, and the system potentially
ends up in a cycle of cooling, clump formation, clump destruction and
heating, so that there is a large amount of molecular gas present in
the outflow at any given time.

\subsection{Outflow visibility}

There are other signatures of AGN outflows in addition to molecular
emission lines. The very high temperature of the wind shock close to
the AGN should make these objects intrinsically luminous gamma-ray
sources. Assuming that the bubble expands spherically with constant
velocity $v_{\rm e}$, we get the shocked wind temperature
\begin{equation}
T_{\rm bub} = T_{\rm sh} \frac{v_{\rm w}^2}{v_{\rm out}^2} \simeq 2
\times 10^{10} {\rm K}.
\end{equation}
At this temperature, a significant fraction of the thermal
bremsstrahlung radiation is emitted as gamma-rays. The number density
in the shocked wind gas, assumed constant throughout the bubble
volume, is
\begin{equation}
n_{\rm bub} = 0.01 \sigma_{200}^2 t_7^{-2} \frac{f_{\rm c}}{f_{\rm
      g}} {\rm cm}^{-3},
\end{equation}
with $t_7 \equiv t/(10 {\rm Myr})$ and assuming that the SMBH mass $M
= M_\sigma$ (eq. \ref{eq:msigma}). The total bremsstrahlung luminosity
emitted in a bubble with volume $V = 4\pi v_{\rm e}^3 t^3/3$ is
\begin{equation}
L_{\rm ff} = 2.4\times 10^{-27}V n_{\rm bub}^2T_{\rm bub}^{1/2} \simeq
3\times 10^{39} \sigma_{200}^6 t_7^{-1} {\rm erg s}^{-1},
\end{equation}
This luminosity is small compared with the total kinetic power of the
outflow $L_{\rm kin} = 0.05 L_{\rm Edd} \simeq 5 \times
10^{44}$~erg/s. This suggests that gamma-ray emission might only be
strong in the early phases of the energy-driven outflow. However, the
strong shocks we predict also offer another route to producing them,
as they can accelerate cosmic ray electrons and produce gamma rays via
inverse Compton or synchrotron emission. In both cases, strong
molecular outflows are potentially rewarding targets for future gamma
ray observatories. The cooling shocked outflow produces thermal
X-rays. Both gamma rays and X-rays are observed in our Galaxy, where
two gamma-ray bubbles disposed symmetrically on either side of the
Galactic plane, surrounded by soft X-ray emission on the edges, were
discovered using the {\it Fermi} satellite \citep{Su2010ApJ}. These
features may be a local (but intrinsically much weaker due to low gas
fraction) example of AGN feedback \citep{Zubovas2011MNRAS}.

We also find that the expected self-regulating star formation rate in
an outflow is large, reaching $\sim 100 \msun$~yr$^{-1}$ or more. The
luminosity of these young stars can outshine the AGN itself, leading
to the galaxy being classed as a LIRG or ULIRG. This misclassification
might be one of the reasons why similar large-scale outflows are
attributed to driving by stellar feedback. Even if the feedback energy
injected by the young stars is larger than the AGN contribution, the
ultimate cause of the whole process is AGN wind.

Another reason for attributing outflow driving to starbursts is the
fact that the outflows we have considered here can persist for $\sim
10^8$~yr after the central AGN has turned off
\citep{King2011MNRAS}. Therefore, outflows should be detected much
more often than the AGN that inflated them in the first place. This
opens the possibility that most, if not all, galaxy-wide outflows are
ultimately triggered by AGN. This seems particularly likely in cases
where the outflow has high velocities, $v_{\rm e} \gtrsim 1000$~km/s,
since starburst-driven winds are unlikely to be capable of reaching
such velocities \citep{Sharma2013ApJ}.

\section{Summary and conclusion} \label{sec:summary}

In this paper, we showed that spherically-symmetric energy-driven AGN
outflows are thermally unstable and rapidly cool down to temperatures
below $10^4$~K, freezing out into dense molecular clumps. This
scenario explains the fact that observed AGN outflows are mostly
composed of molecular gas moving at high velocities. These findings,
together with previous work analysing the structure of such outflows,
give the following predictions for outflow visibility:

\begin{itemize}

\item The shocked wind bubble emits gamma rays. Unfortunately, the
  expected luminosity of the whole bubble ($L_\gamma \leq
  10^{39}$~erg/s) is too low to be detectable with present
  instruments.

\item The hot diffuse outflowing gas creates a soft X-ray envelope
  around the gamma-ray bubble. Parts of this envelope may be visible
  close to galactic discs, where the ISM density is much larger than
  elsewhere in the bulge and halo.

\item The cold clumps, which account for most of the outflow mass, are
  observable in molecular lines. They are unlikely to form in galaxies
  with depleted diffuse gas content, such as the Milky Way, but should
  be common in dense systems.

\item Star formation in the molecular clumps proceeds at a rate of
  order $100 \; \msun$~yr$^{-1}$, enough for its luminosity to exceed
  the total luminosity of the AGN.

\item Many outflows attributed to star formation may instead be
  energy-driven by AGN. A direct signature of this would be a central
  gamma-ray emitting bubble surrounded by an X-ray emitting
  shell. Since the outflow must be energy-driven this in turn would
  imply a central black hole mass very close to the $M - \sigma$
  value. Such AGN--driven outflows should be both faster ($v_e \ga
  1000$~km s$^{-1}$) and more concentrated to the galaxy centre than
  those driven by starbursts.
\end{itemize}

\section*{Acknowledgments}

We thank Roberto Maiolino and Sergei Nayakshin for helpful
discussions, suggestions and comments on the manuscript. We thank the
referee, Biman Nath, for a very helpful report. KZ thanks the UK STFC
for support in the form succesively of a postgraduate studentship and
a postdoctoral research associate position, both at the University of
Leicester. KZ is funded by the Research Council of Lithuania grant
no. MIP-062/2013.

\end{document}